\documentclass[letters,useAMS,usenatbib]{mn2e}
\usepackage{graphics}
\newcommand{\um}{\,$\mu$m}
\newcommand{\cyga}{Cygnus~A}
\newcommand{\kms}{\,km\,s$^{-1}$}
\newcommand{\dgr}{$^\circ$}

\title[Polarization and kinematics in Cygnus~A]
{Polarization and kinematics in Cygnus~A}

\author[van Bemmel et al.]{Ilse M. van Bemmel$^1$\thanks{ESA fellow, on assignment from Space Telescope Division of the European Space Agency}
, Jo\"el Vernet$^2$, 
Robert A.E. Fosbury$^3$, Henny J.G.L.M. Lamers$^4$\\
\\
$^1$ Space Telescope Science Institute, 3700 San Martin Drive, Baltimore, 
MD 21218, USA, bemmel@stsci.edu\\
$^2$ Osservatorio Astrofisico di Arcetri, Largo E. Fermi 5, I--50125 Firenze,
Italy\\
$^3$ Space Telescope European Coordinating Facility, Karl-Schwarzschild-Strasse 2, D--85748 Garching, Germany\\
$^4$ Sterrenkundig Instituut Utrecht, P.O.Box 80000, NL--3508 TA Utrecht,
The Netherlands
}
\begin{document}

\date{Submitted June 3, 2003}

\pagerange{\pageref{firstpage}--\pageref{lastpage}} \pubyear{2003}


\maketitle

\label{firstpage}

\begin{abstract}
From optical spectropolarimetry of \cyga\ we conclude that the
scattering medium in the ionization cones in \cyga\ is moving outward
at a speed of 170$\pm$34\kms, and that the required momentum can be
supplied by the radiation pressure of an average quasar. Such a
process could produce a structure resembling the observed ionization
cones, which are thought to result from shadowing by a circumnuclear
dust torus.  We detect a polarized red wing in the [O\,III] emission
lines arising from the central kiloparsec of \cyga. This wing is
consistent with line emission created close to the boundary of the
broad-line region.
\end{abstract}


\section{Introduction}
Dust in active galaxies provides a screen that can both obscure the
central regions from direct view, but also scatter nuclear radiation
into our line of sight. By studying the scattering-induced
polarization of the emission lines, some aspects of the kinematics of
the scattering medium can be deduced \citep{wal94}.  In this paper we
study the polarization and kinematics of line emitting regions in the
radio galaxy \cyga .

\cyga\ is one of the brightest and nearest powerful double-lobed radio
galaxies. Its proximity allows us to spatially resolve the two
narrow-line regions on either side of the nucleus with sufficient
signal-to-noise to study the polarization of the emission lines.
Hence, it is an ideal object to study the physics of the narrow-line
region. From radio observations we know that the western side is
oriented towards us, with the radio jet inclined at about 50$^{\circ}$
to our line of sight \citep{car96}.  We will refer to the western
side as the near side, and to the eastern side as the far side.

The optical continuum polarization has been studied in detail by
\citet[][hereafter O97]{ogle97}. They find a strong blue, polarized
continuum in the west, consistent with the presence of a medium power
central quasar. In the east the emission is less polarized, which they
ascribe to dilution by a young stellar population. Both in direct and
in polarized light they detect a broad component underlying the narrow
H$\alpha$ emission line. In the west they also detect polarized broad
H$\beta$, in agreement with the orientation of \cyga, and confirming
the presence of a hidden quasar.

Near- and mid-infrared images show biconical structures, resembling
the optical ionization cones \citep{jack98}.  In the near-infrared,
the edges of the cones are bright and highly polarized \citep{tadh99},
indicating that the walls of the ionization cones are dusty, and
illuminated by the central quasar. In a later paper \citep{tadh00}, it
is shown that polarization is only seen along one side of the
bicone. No explanation has yet been found for this
asymmetry. Mid-infrared imaging at 10 and 18\um\ confirms the
existence of a conical shape, which is especially prominent on the far
side \citep{rad02}.

Ionization cones are common in active galaxies. The two physical 
processes that can generate low density cones are
obscuration close to the central source as in young stellar objects,
and suspected to be present in most active galaxies, or a bi-polar 
outflow as detected in planetary nebulae.
Obscuration by a dusty torus is postulated as part of the unification
scheme for radio-loud active galaxies \citep{up95}. However, outflow
of matter has been detected in broad absorption line quasars, where
gas moving away from the nucleus causes broad absorption lines in 
the optical and UV spectrum. \cyga\ shows outflowing gas at larger
distance which emits [OIII] lines \citep{tadh94}.
Outflows can excavate funnels
in the galaxy, in which the ISM will be ionized by the central
source. They can be driven by either the radiation pressure of the
central quasar, or a nuclear starburst and do not necessarily require
any obscuration to produce sharp shadows. Using existing deep optical
spectropolarimetry of \cyga, we can determine aspects of the
kinematics of the narrow-line region, and infer the process that
dominates the creation of the ionization cones.

For the remainder of this paper we will use 
$H_0 =$65\,km\,s$^{-1}$\,Mpc$^{-1}$ and $q_0 =$0.5, and
a redshift for \cyga\ of $z=0.0561$, determined from
the redshift of the nuclear [OIII] doublet in our data.

\section{The data reduction}
For our analysis we will use the same data as used in the O97
paper. They
consist of deep spectropolarimetry of Cygnus~A, obtained with the
Low Resolution Imaging Spectrograph polarimeter \citep{oke95,good95} at
the Keck telescope during the night of UT 1996 October 4--5. A longslit of
1$''$ width was used, aligned with the radio axis at PA=101$^{\circ}$. 
Two sets of data were obtained, one in good seeing (0.6$''$) of
$4\times900$ seconds, and a second with worse seeing (1.0$''$) of
$4\times1200$ seconds.  Both sets contain observations at four different 
angles of the half-wave plate (HWP), i.e. 0\dgr, 22.5\dgr, 45\dgr\ and 
67.5\dgr. The spectra cover 3500 to 8500\,\AA\ in the observed frame,
with a dispersion of 2.5\,\AA\ per pixel and effective resolution of 
10\,\AA.

Data reduction on the spectra was performed separately for each 
orientation of the HWP, splitting the usual ordinary and extra-ordinary 
rays. Standard IRAF routines were used for bias subtraction, flat
field division, wavelength calibration from NeAr arc spectra, and
geometrical correction.  The final 3$\sigma$ error in both the
wavelength calibration and the geometrical correction is 0.6\,\AA,
which corresponds to one fourth of the spectral pixel size.  Three
0.85$''$ apertures are traced in all separate 2-dimensional spectra,
located 1.2\,kpc east of the nucleus, at the 
nucleus itself, and at 1.2\,kpc west of the nucleus. The width of the
apertures corresponds to 1\,kpc in the host galaxy. The apertures used 
for sky
subtraction are 10$''$ from the nuclear emission region on either side. 
The two sky apertures are averaged before subtraction.
After the complete reduction the two data sets with different seeing are
combined to obtain a better signal to noise in the emission lines. No
galaxy subtraction is performed.

After suitable binning, the 8 rays are combined to form the Stokes
parameters $I$, $Q$ and $U$, from which we compute the fractional
polarization. We have corrected for the Ricean bias by using a Monte 
Carlo technique \citep{fos93}. 
This method provides a distribution function of $U$, $Q$,
$P$ and $\theta_P$, and gives an estimate of their errors. The complete 
technique is
described in detail by \citet{ver01}. The resulting value of
$\theta_P$ is corrected for the instrumental offset and calibrated using 
observations of the polarized standard star HD\,155528.

\begin{figure}
\begin{center}
  \resizebox{9.cm}{!}{\hspace{-10mm}\includegraphics{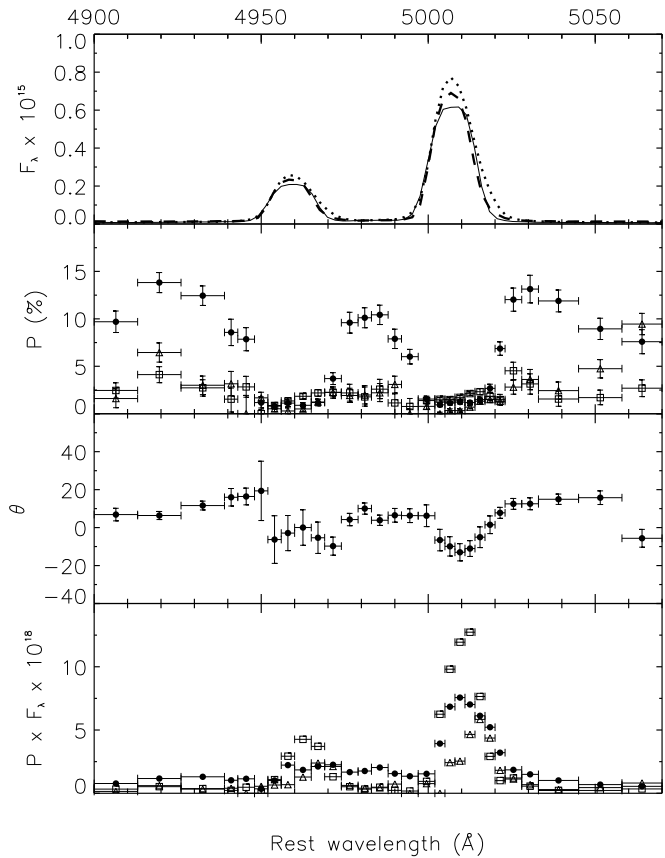}} 
  \caption{\label{line_pol} Polarization in the [O\,III]
  doublet. From top to bottom is shown the total flux in
  erg\,cm$^{-2}$\,s$^{-1}$\,\AA$^{-1}$, fractional polarization, the
  polarization angle of the E-vector in the western aperture and the
  polarized flux in erg\,cm$^{-2}$\,s$^{-1}$\,\AA$^{-1}$. In the total
  flux spectrum the apertures are: west=solid line, nucleus= dotted
  line, east=dashed line. In the remaining three plots the apertures
  are coded as follows: triangles=east, squares=nucleus, filled
  circles=west. The polarization angle is only plotted for the western
  aperture, the trend is similar in the other apertures.  }
\vspace{-5mm}
\end{center}
\end{figure}

\section{Results}
In this paper we will focus on the polarization of the emission lines.
For the results and analysis of the continuum polarization and
its different spectral components, we refer to the O97
paper. The strongest narrow lines were selected to calculate the fractional
polarization, polarization angle and polarized flux over the profile. The
general observed trend in all these lines is well illustrated by the 
[O\,III] doublet at 4959,5007\,\AA, shown in Fig.~\ref{line_pol}.

\begin{figure}
\begin{center}
  \resizebox{9.cm}{!}{\hspace{-10mm}\includegraphics{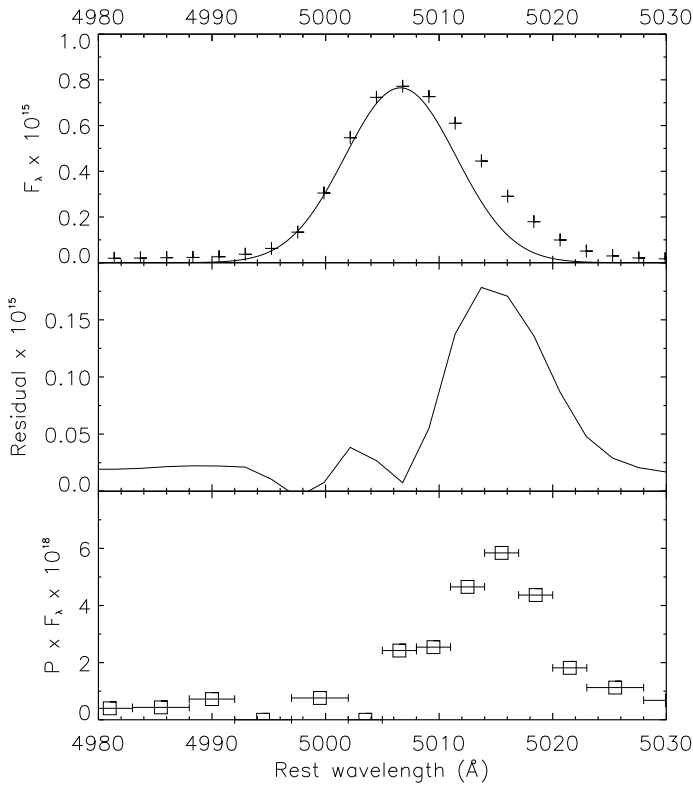}} 
  \caption{\label{oiii_wing} Demonstration of the polarized wing on
  the nuclear [O\,III]5007\,\AA\ line. The top panel shows a best fit 
  Gaussian with fixed central wavelength
  (line) to the observations (pluses), and in the middle panel the
  residual of the fit is shown. The bottom panel displays the
  polarized flux in the line. Units as in Fig.~\ref{line_pol}.}
\end{center}
\end{figure}

We observe significant polarization of $P=2\pm1$\% in the emission 
lines in the east and west. The fractional polarization decreases 
from the continuum value towards the line center.
In the nucleus $P=1\pm1$\%, consistent with no polarization. The value 
of $P=2\pm1$\% is consistent with the value found by O97 of $\sim1.3$\%.
In all apertures we observe a peak in polarization 
$\sim20$\,\AA\ redward of the line center. In the east and nucleus
this causes a redshift of a few \AA ngstr\"oms in the polarized line 
flux. In the western aperture a rotation in polarization angle
of $\sim20^{\circ}$ is observed over the line. The same trend is seen 
in the other apertures, albeit with larger errors.

The wavelength shift of the polarized line is particularly prominent 
in the nucleus, where the polarized line corresponds to a red wing 
observed in the total flux line. This wing, as shown in Fig.\ref{oiii_wing},
is not seen in the other apertures, and remains 
undetected in other strong emission lines. However, these are all 
significantly weaker, or blended in the red. Hence, the presence 
of a red wing as seen in the nuclear [O\,III] doublet cannot be 
excluded. 

Deviations in the fractional polarization can be induced during the
data-reduction process, especially by residual wavelength calibration
errors between the o- and e-rays. To assess whether the observed red 
peak in the polarization is due to such an error, we constructed a simple 
model that reproduces the effects of line broadening and wavelength 
shifts in a single ray
on the resulting polarization of a Gaussian emission line. The
only way to reproduce a feature as observed in \cyga, is by changing
the line width and offsetting the line center in two of the ordinary
ray frames. This is impossible to reconcile with the fact that all 
data are reduced identically. In addition, the offset needs to be 10 
times larger than the achieved wavelength calibration error. Therefore, 
we conclude that the observed signature is intrinsic.

\section{Analysis of the line polarization}
As discussed above, the fractional polarization of the emission lines
shows a peak redward of the line center. Similar behaviour has been
observed in the planetary nebula NGC\,7027 \citep{wal94}, where it
is interpreted as an outward movement of the scattering medium. If
the central quasar in \cyga\ can provide enough radiation pressure 
to push the matter in the ionization cones outward, this implies 
a similar situation can occur.

Treating scattering as instantaneous absorption and re-emission,
it is easy to see that scattering of photons by a particle
moving away radially from the source of emission will produce
a Doppler shift. The observer sees the photon with a wavelength
$\lambda_{obs}$, given by:
\begin{displaymath}
\lambda_{obs}=\lambda_0 [1 +\frac{v}{c}(1-\cos{\theta})],
\end{displaymath}
in which $\theta$ is the angle between the direction of motion
of the scattering particle and the line of sight to the observer,
$v$ is the particle velocity, $\lambda_0$ is the wavelength
at which the photon is emitted, and assuming $v \ll c$.

\begin{figure}
\begin{center}
  \resizebox{8.7cm}{!}{\includegraphics{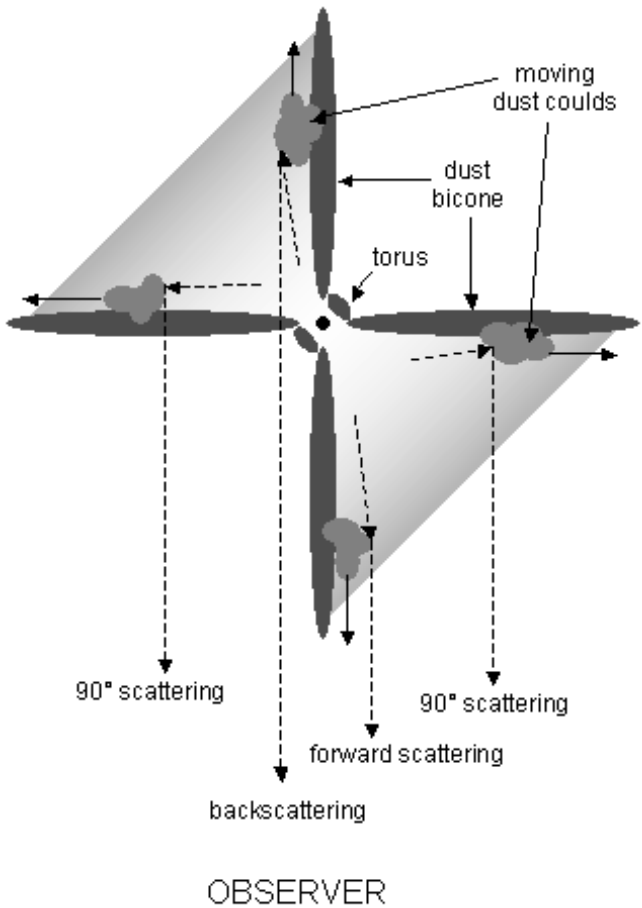}}
  \caption{\label{scat_geom} Cartoon of the scattering geometry
  in \cyga. The solid arrows note the direction of the velocity of the
  dust clouds, the dashed arrows the direction of the light. East is
	to the left, west to the right, to observer is down.}
\end{center}
\end{figure}

The effect of the moving dust on the resulting polarized spectrum
depends only on the angle of observation, $\theta$.  The situation is
illustrated in Fig.~\ref{scat_geom}.  The term $1+\frac{v}{c}(1-\cos{\theta})$
reaches a maximum of $1+ \frac{2v}{c}$ for backward scattering. To
the observer, this
occurs only on the eastern side of the galaxy, so here a maximum shift
is expected in the polarized line flux. For scattering through
90$^{\circ}$ the term $1-\cos{\theta}$ equals 1, which occurs on
both sides.  Therefore, it cannot generate an observed difference
between the polarized emission lines from either side. In the case of
forward scattering, the cosine term cancels, and the wavelength of the
scattered emission is equal to the emitted wavelength. This is only
possible on the western side. For scattering through angles close to 0
or 180$^{\circ}$ there is little polarization. However, the scattered
emission on the both sides arises from an integral over scattering
angles between 90$^{\circ}$ and either 0$^{\circ}$ or 180$^{\circ}$, 
which can result in the 3\% polarization in the emission lines.

By measuring the offset of the peak of the line in polarized flux on
the east, the expansion velocity of the scattering medium can be
obtained. From radio observations we know that the angle of the jet is
$\sim$50$^{\circ}$ \citep{car96}, which corresponds to an angle of 
$\sim$140$^{\circ}$ between the direction of motion of the scattering
medium and our line of sight. The offset is measured to be
5\,$\pm$\,1\,\AA, which yields an outflow velocity of
170\,$\pm$\,34\kms.  This is very small compared to outflow velocities
of several to tens of thousands \kms\ that are generally measured in 
broad-absorption line quasars \citep{elvis00}, and
suggests that the scattering medium is unrelated to the high velocity
clouds that produce the UV and X-ray absorption lines in such
systems. \\

To understand the nature of the driving force, we will estimate the
expected velocity of dust clouds driven by the radiation pressure from
the central source.  A grain will be pushed outward if it experiences
a radiation pressure larger than the force of gravity. The
radiative acceleration on a single grain is given by:
\begin{equation}
g_{\rm rad} = \frac{\pi a^2 Q L}{4 \pi r^2 c\ m_d}
\end{equation}
\citep[see][p.165]{lam99}, in which $Q$ is the radiation
pressure efficiency, which we will assume to be unity, $a$ the radius
of the grain, $L$ the total luminosity of the central source, $r$ the
distance, $c$ the speed of light and $m_d$ the mass of a single
grain. The grain mass is given by $m_d = \frac{4}{3} \pi a^3 \rho_s$,
and the specific density of the grain material $\rho_s
=3.6$\,g\,cm$^{-3}$ for silicate grains and 1.8\,g\,cm$^{-3}$ for
carbonaceous grains. 
Classic interstellar dust grains extend a range of sizes from 0.001\um\
to 10\um\ \citep{laor93}. We assume an average grain radius of 0.1\um\
as in \citet{rad02}, to balance the effects of larger acceleration for
smaller grains, and lower abundance of the larger grains.
We also assume pure silicate dust, adding carbonaceous grains to 
the dust will increase the radiative acceleration.
Furthermore, we adopt a nuclear luminosity $L=10^{45.4}$\,erg\,s$^{-1}$
based on the recent X-ray observations by \cite{young02}. At
a distance of 1\,kpc, the
radiative acceleration in \cyga\ equals 5.1$\times10^{-6}$\,cm\,s$^{-2}$. 
The gravity acceleration exerted on the grain depends on the mass 
enclosed within the radius at which the grain sits:
\begin{displaymath}
g_{\rm g} = \frac{G M}{r^2} = M_r({\rm M_{\odot}}) \cdot
1.4\times10^{-17} {\rm cm\,s^2}
\end{displaymath}
assuming again a distance of 1\,kpc, and $M_r$ being the mass within
that radius in units of solar mass.  The radiative acceleration exceeds the
gravity if $M_r < 3\times10^{11}$\,M$_{\odot}$.  This is equivalent to 
the mass
of a large spiral galaxy, and about two orders of magnitude larger
than the mass of the super-massive black hole in \cyga\ of 
$\sim2\times10^9$\,M$_{\odot}$ \citep{tadh03}, which is expected to dominate 
the central potential.  Thus we can safely assume that the
dust will be accelerated by radiation pressure.

With $M_r \ll 10^{11}$\,M$_{\odot}$ we can express the acceleration
of a dust cloud in terms of the central luminosity, dust-to-gas
ratio ($\delta$) and distance from the central source:
\begin{displaymath}
g_{\rm cloud} = \frac{N_d \pi a^2}{M_{\rm cloud}} \cdot
\frac{Q L}{4 \pi r^2 c} = \frac {C}{r^2},
\end{displaymath}
where $N_d$ is the total number of dust grains in an optically
thin cloud, given by
\begin{displaymath}
N_d = \frac{3 \delta M_{\rm cloud}}{4 \pi a^3 \rho_s}
\end{displaymath}
and 
\begin{displaymath}
C = \frac{3 \delta Q L}{16 \pi a \rho_s c} = 1.4 \times 10^{36}
\end{displaymath}
in cgs units, for the values of the parameters adopted above.
We use a Galactic dust-to-gas ratio of $\delta=0.01$.

The solution of the equation of motion gives the cloud velocity
as a function of distance to the central source:
\begin{displaymath}
v(r) = v_{\infty} \sqrt{1-\frac{r_0}{r}},
\end{displaymath}
where $r_0$ is the distance from the nucleus where the dust cloud is
formed, and the terminal velocity $v_{\infty}$ is given by
$\sqrt{2C/r_0}$. Adopting a cloud velocity of 170 km/s at a distance
of 1\,kpc, we find that the cloud must have formed at $r_0 =
0.8$\,kpc. This implies that the scattering medium originates within
the 1\,kpc aperture. If it had been created in the nuclear regions 
and accelerated from there, its velocity
at 1\,kpc would have been much higher than observed. Presumably the
dust from the dense layer at the edge of the bicone is entrained in
the turbulent flow in the bicone. When it enters the radiation field
of the central source, it will experience the radiation pressure. The
dust clouds will then move away from the nucleus, while scattering the
light from the central source.  
This is consistent with the hypothesis
derived from the infrared appearance of the bicones that their central 
regions have been cleared from dust by radiation pressure, as
proposed by \cite{tadh00}.

Two basic assumptions are made. First of all, we assume the
scattering cloud to be optically thin. If the cloud were
optically thick, it would have to be created closer to the
central source, since its effective scattering surface will
be smaller, and the coupling of radiation to the dust particles
will be less efficient. The second basic assumption is that
the drift velocity is small compared to the thermal speed of
the gas, which implies that all momentum
that is transfered from radiation to the dust cloud is shared
by all dust and gas particles via internal collisions. This
assumption is valid if the gas density is above a critical
value that can be derived from eq.\,7.44 in \citet{lam99}
by setting the drift velocity equal to the thermal speed.
Adopting a temperature of $10^4$\,K \citep{ost78} the thermal
speed of the gas is of order 10\kms. This results in a lower
limit for the electron density of the gas of 40\,cm$^{-3}$.
The typical observed electron density in narrow-line regions 
lies around $10^{3-4}$\,cm$^{-3}$ \citep{ost78,fer81}, several
orders of magnitude above the critical lower limit, so this
assumption is valid.

The use of polarization shows unambiguously that an outflow must be
present in the ionization cones. In case of an infall of matter, the
absorbed emission will have a blueshift:
$\lambda_{obs} = \lambda_0[1-\frac{v}{c}(1+\cos\theta)]$.  
On the side where $\theta$ is small a blueshift of the
line in polarized flux is expected, while for large $\theta$ no net
effect should be visible. In the geometry of \cyga\ this would correspond
to a blueshift in the eastern aperture.  Using polarimetry of
forbidden line profiles thus provides a unique method to measure
infall and outflow velocities in any astronomical object. 

\section{Conclusions}
Spectropolarimetry of (active) galaxies provides a powerful probe 
of the kinematics of the different components present. It can be 
used to uniquely determine outflow or infall of matter in any 
sufficiently polarized astronomical object.

Large scale mass outflow occurs in Cygnus~A, driven by the 
radiation pressure of the central source. The outflow velocity is 
measured to be $170\pm34$\,km\,s$^{-1}$. This causes
changes in the fractional polarization over the narrow emission
lines, showing a shift of the line peak in polarized flux to the 
red. The polarized line flux from
the east is redshifted, while on the west no shift is observed.
This is only consistent with an outflow, not with infall of matter.

A previously undetected redshifted nuclear component is found in
[O\,III], which is highly polarized. This component is consistent with
line emission generated very close to the broad-line region, which is
obscured from direct view. It is consistent with previous observations 
of anisotropic [O\,III] emission {\citep{hes93}, and requires an
obscuring medium of some sort in our line of sight to the inner
narrow-line region. This nuclear component
is not observed in any other narrow line, but its presence cannot be
excluded due to blending or lower signal-to-noise in the other
lines.

The detection of an outward moving scattering medium in the ionization 
cones suggests that their shape could be determined by an outflowing
wind driven by the radiation from the central quasar coupling to the
dust. This is consistent with previous results by 
\citet{tadh99,tadh00}. As a consequence, the detection of ionization 
cones in active galaxies cannot necessarily be used as evidence in favour of 
orientation-based unification, although it does not exclude the presence 
of an obscuring 
torus. The high obscuration of the central regions in \cyga\ and other radio
galaxies could also be attributed to large scale dust. Detailed near- and 
mid-infrared observations, and modeling the emission of obscuring 
tori are required to give direct evidence of their presence.

{\it Acknowledgements} 
We wish to thank Pat Ogle, Hien Tran, Joe Miller,
Bob Goodrich and Marshall Cohen for initial involvement. Furthermore,
thanks to Xander Tielens and Peter Barthel for advise and critical
comments. Thanks to Jeremy Walsh for initial involvement. And thanks 
to the anonymous referee for suggesting small but significant
refinements of the text. IMvB wishes to thank ESA, STScI DDRF
grant 41559, ESO and the Kapteyn Institute, Groningen, the Netherlands, 
for financial support. RAEF is affiliated  with the Space Telescope 
Division of the European Space Agency, ESTEC, Noordwijk, the Netherlands.

\bibliographystyle{/data/brolga6/bemmel/Tex/BibTex/mn2e}
\bibliography{/data/brolga6/bemmel/Tex/BibTex/ivb}

\label{lastpage}

\end{document}